\documentclass[12pt]{iopart}

\usepackage{graphicx}
\usepackage{setstack}
\usepackage{iopams}

\usepackage{float}
\restylefloat{table}
\begin{document}
\title{Synthesis, structure and magnetism of the new $S=\frac{1}{2}$ 
kagome magnet NH$_4$Cu$_{2.5}$V$_2$O$_7$(OH)$_2$.H$_2$O}

\author{E Connolly$^1$, P Reeves$^{1,2}$, 
D Boldrin$^3$ and A S Wills$^1$}

\address{$^1$ Department of Chemistry, University College London, 20 Gordon St, London WC1H 0AJ}
\address{$^2$ Present address: Department of Material Science and Metallurgy, University of Cambridge, 27 Charles Babbage Road, Cambridge CB3 0FS, UK}
\address{$^3$ Department of Physics, 
Imperial College, Prince Consort Road, London SW7~2BZ, UK}
\ead{a.s.wills@ucl.ac.uk}

\begin{abstract}
The study of quantum spin-liquid states (QSL) with lattice dimension $>1$ has proven an enduring problem in solid state physics. Key candidate materials are the $S=\frac{1}{2}$ kagome magnets due to their ability to host quantum fluctuations within the high degeneracy of their frustrated geometries.  Studies of an increasing library of known $S=\frac{1}{2}$ kagome magnetic materials has challenged our understanding of the possible QSL states, for example, the recent discovery of a chiral spin-liquid ground state in kapellasite showed that even magnets with ferromagnetic nearest-neighbour exchange are not necessarily trivial and that QSL states beyond the superposition of simple singlet are possible.

Here, we outline the synthesis, structure and preliminary magnetic characterisation of a candidate QSL material, the $S=\frac{1}{2}$ kagome magnet NH$_4$Cu$_{2.5}$V$_2$O$_7$(OH)$_2$.H$_2$O. The crystal structure of NH$_4$Cu$_{2.5}$V$_2$O$_7$(OH)$_2$.H$_2$O has the 3-fold symmetry of a geometrically `perfect' kagome lattice while the magnetism shows a competition between ferromagnetic and antiferromagnetic characters reminiscent of kapellasite. 
\end{abstract}
\submitto{\JPCM}

\section{Introduction}
\label{Int}

\begin{figure*}[t]
\includegraphics[scale=0.15]{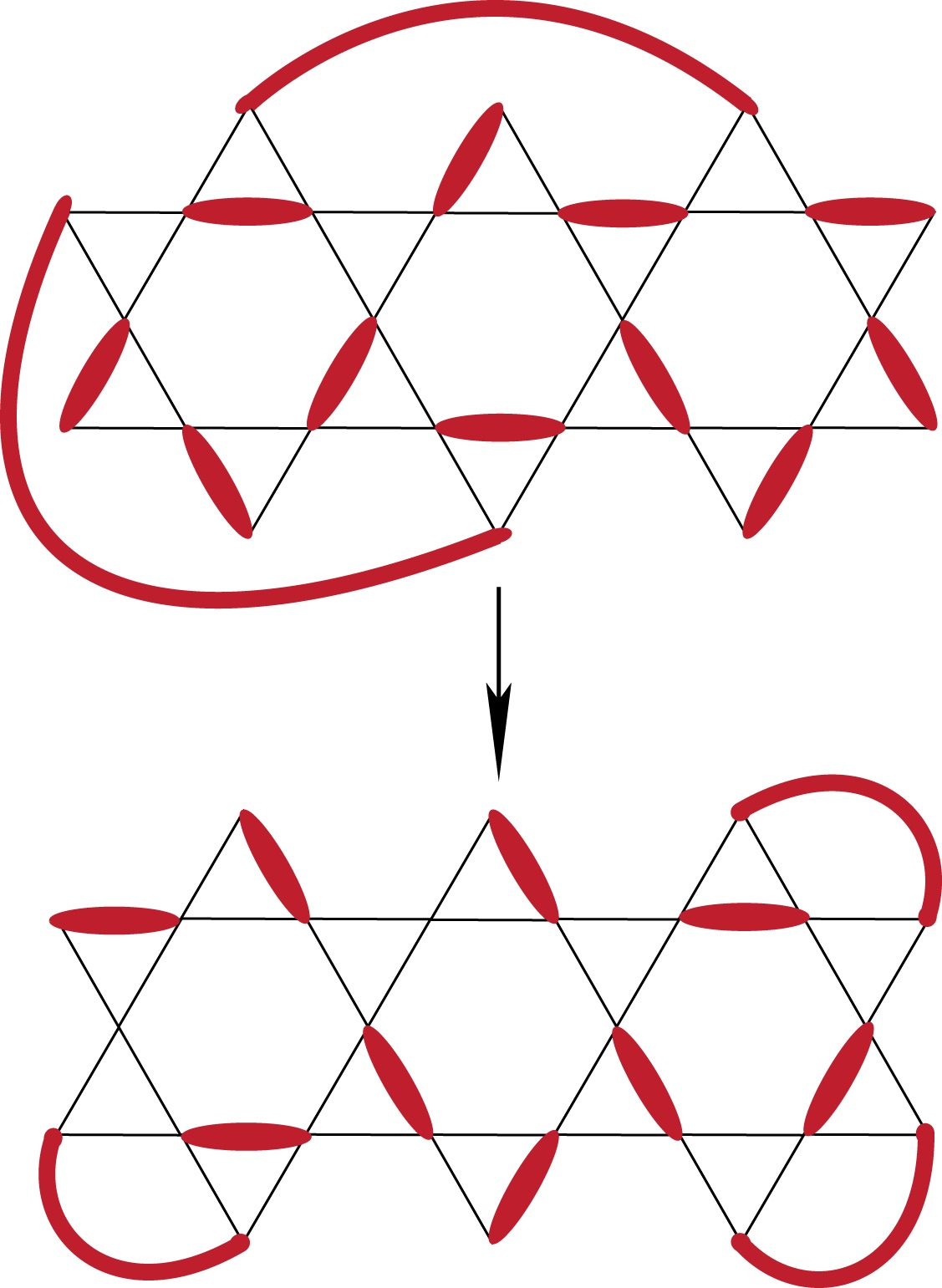}\centering
\caption{Schematic of a kagome structure hosting a QSL state. The short and long range entangled spins are highlighted in red. The entangled singlet states are in a disordered arrangement over the kagome lattice. The degeneracy of the ground state allows for zero-point energy fluctuations to other disordered arrangements.}
\label{QSL}
\end{figure*}

The quest to discover what ground states occur when quantum fluctuations destabilize conventional magnetic order has become one of the backbones of modern condensed matter physics. Much of this work has focused on $S=\frac{1}{2}$ kagome magnets, where the moment-bearing ions make up a 2-dimensional network of vertex-sharing triangles. Unlike the ground states of square and triangular $S=\frac{1}{2}$ Heisenberg magnets, the $S=\frac{1}{2}$ kagome Heisenberg antiferromagnet was shown theoretically not to order into a N\'eel state, even at $T=0$~K\cite {Rokhsar1988, Sachdev1992, Lecheminant1997}. Research into these quantum frustrated magnets has largely followed the picture of the ground states being dynamic superposition states of degenerate local- or long-range entangled singlet pairs (Figure~\ref{QSL}), quantum spin liquids (QSLs). QSLs differ fundamentally from conventional magnetic order in that they do not break translational or rotational symmetry, but are instead defined by topological order parameters \cite{Balents2010}.

Recent experimental and theoretical investigations into $S=\frac{1}{2}$ kagome magnets have revealed that new types of QSL states are able to form when nearest-neighbour ferromagnetic exchange is frustrated by antiferromagnetic further-neighbour interactions\cite{Fak2012, Boldrin2015a}, thereby expanding the field from one that was thought to be only relevant for nearest-neighbour  antiferromagnets to those with competing signs of exchange. This work, focused on kapellasite ( $\mathrm{\alpha-ZnCu_3(OH)_6Cl_2}$) - which has a 12-sublattice chiral spin-liquid (cuboc2) ground state \cite{Fak2012} - and its isostructural and isomagnetic analogue haydeeite ($\mathrm{\alpha-MgCu_3(OH)_6Cl_2}$) helped formulate a new phase diagram for the QSL ground state involving a diagonal exchange integral. Changes in its value caused by differences in the bonding around their diamagnetic ions (Mg$^{2+}$ and Zn$^{2+}$), is able to drive the conventional ferromagnetic ordering seen in haydeeite into the 12-sublattice chiral spin-liquid (cuboc2) ground state\cite{Fak2012, Boldrin2015a, Colman2008, Janson2008, Colman2010}.

At present, much of the research into experimental $S=\frac{1}{2}$ kagome magnets is based on two main families of crystal structures: the  atacamites (herbertsmithite $\mathrm{\gamma-ZnCu_3(OH)_6Cl_2}$, kapellasite $\mathrm{\alpha-ZnCu_3(OH)_6Cl_2}$ and their isomagnetic relatives `Mg-herbertsmithite' $\mathrm{\gamma-MgCu_3(OH)_6Cl_2}$, and haydeeite $\mathrm{\alpha-MgCu_3(OH)_6Cl_2}$) and the copper vanadates (volborthite $\mathrm{\alpha-Cu_3V_2O_7(OH).2H_2O}$, vesignieite $\mathrm{BaCu_3V_2O_8(OH)_2}$ and `Sr-vesignieite' $\mathrm{SrCu_3V_2O_8(OH)_2}$). The differences between these magnets has been revealing. The atacamites all have 3-fold symmetry, a quality that brings them close to the theoretical models.  In the well studied herbertsmithite, the QSL state survives Dzyaloshinskii-Moriya (DM) anisotropy as the anisotropic exchange is largely axial and its strength is below a quantum critical point, $D_z^C/J \simeq 0.1$. In contrast, the  DM component in vesignieite is dominated by the in-plane component, $D_p$ \cite{Zorko2008, Zorko2013}, despite $D_z/J$ being similar to  herbertsmithite, and this induces partial ferromagnetic order\cite{Zorko2013}. Even closely related materials can show wildly different behaviours, such as the QSL of dynamic singlets of herbertsmithite and of cuboc2 spin correlations in its polymorph kapellasite. At the root of this richness of properties are often subtleties in the crystal structure, such as variation in bond angles related to site disorder or differences in orbital ordering patterns and consequent superexchange pathways, as seen in volborthite\cite{Hiroi2001}, vesignieite\cite{Okamoto2009a,Colman2011} and `Sr-vesignieite'\cite{Boldrin2015a}.

Here, we present the synthesis, structure and preliminary magnetic measurements of a new $S=\frac{1}{2}$ kagome material NH$_4$Cu$_{2.5}$V$_2$O$_7$(OH)$_2$.H$_2$O based on the copper vanadates, and introduce its magnetic properties.

\section{Synthesis}
\label{Synthesis}

Syntheses using a scaled down version (to 30~\% of the literature quantities) of that previously given in a paper by Palacio\cite{Palacio2008a} at $T=170 ^\circ$C, produced a product contaminated by an  amorphous impurity phase and crystalline $\mathrm{NH_4VO_3}$. The latter could be identified by a peak in the diffraction data at $2\theta=24^\circ$ that is also visible in \cite{Palacio2008a}.  

Further studies showed that the impurities form at $T\leq 80\,^\circ \mathrm{C}$ and $T\geq 130\,^\circ \mathrm{C}$. A phase pure sample was obtained as follows: $\mathrm{NH_4OH}$ (0.62 ml, 32 $\%$ wt, Sigma-Aldrich) was diluted with distilled water (3.88 ml). To this solution, $\mathrm{V_2O_5}$ (166 mg, 99.6 $\%$, Aldrich) was added and the suspension was stirred for 1 hour, whereupon it turned yellow. Finally, $\mathrm{CuCl_2.2H_2O}$ (311 mg, 99.8 $\%$, Aldrich) dissolved in distilled water (3.0 ml), was added to the yellow suspension. This was then stirred for 1 hour to homogenize, producing a turquoise gel. The gel was loaded into a Pyrex pressure tube (15$\,\mathrm{ml}$, Ace Glass Inc.) and then suspended in a silicon oil bath at $115\, ^\circ \mathrm{C}$ for 24 hours. The product was washed 3 times in water $via$ centrifugation ($4.5\times10^3$\,rpm, 2\,mins) and dried at $60 \,^\circ \mathrm{C}$ for 5 hours. A yellow powder of NH$_4$Cu$_{2.5}$V$_2$O$_7$(OH)$_2$.H$_2$O was produced in a yield of $\sim44\,\%$.

The ratios of the final reagent used are 1~V$_2$O$_5$ : 2~CuCl$_2$.2H$_2$O : 6~NH$_4$OH : 422~H$_2$O. Powder XRD indicated that the product of the first reaction is $\mathrm{NH_4VO_3}$. The pH of the second reaction was 9.6 before and after the synthesis. We propose  the following reaction mechanism for the formation of NH$_4$Cu$_{2.5}$V$_2$O$_7$(OH)$_2$.H$_2$O:

\begin{eqnarray*} \label{1}
\mathrm{V_2O_{5(s)} + 2NH_4OH_{(aq)} \rightarrow 2NH_4VO_{3(s)} + H_2O_{(l)}} \\ 
\mathrm{2NH_4VO_{3(s)}  + \frac{5}{2}CuCl_2.2H_2O _{(aq)}} +\,  \mathrm{4NH_4OH_{(aq)}} \rightarrow  \mathrm{NH_4Cu_{2.5}V_2O_7(OH)_2.H_2O_{(s)}} \\ + \mathrm{5NH_4Cl_{(aq)} + 3H_2O_{(l)} + \frac{3}{2}O_{2(g)}} 
\end{eqnarray*}

The hypothesized reaction scheme has a calculated reagent ratio for 1~M of NH$_4$Cu$_{2.5}$V$_2$O$_7$(OH)$_2$.H$_2$O of 1~V$_2$O$_5$ : 2.5~CuCl$_2$.2H$_2$O : 6~NH$_4$OH, indicating that our synthetic conditions correspond to a slight deficit of CuCl$_2$.2H$_2$O. The outlined synthetic procedure can be used to understand how variation to concentrations or reagent types will effect the product.   

\section{Crystal structure determination}
\label{Crystal structure determination}

\begin{figure*}[t]
\includegraphics[scale=0.25]{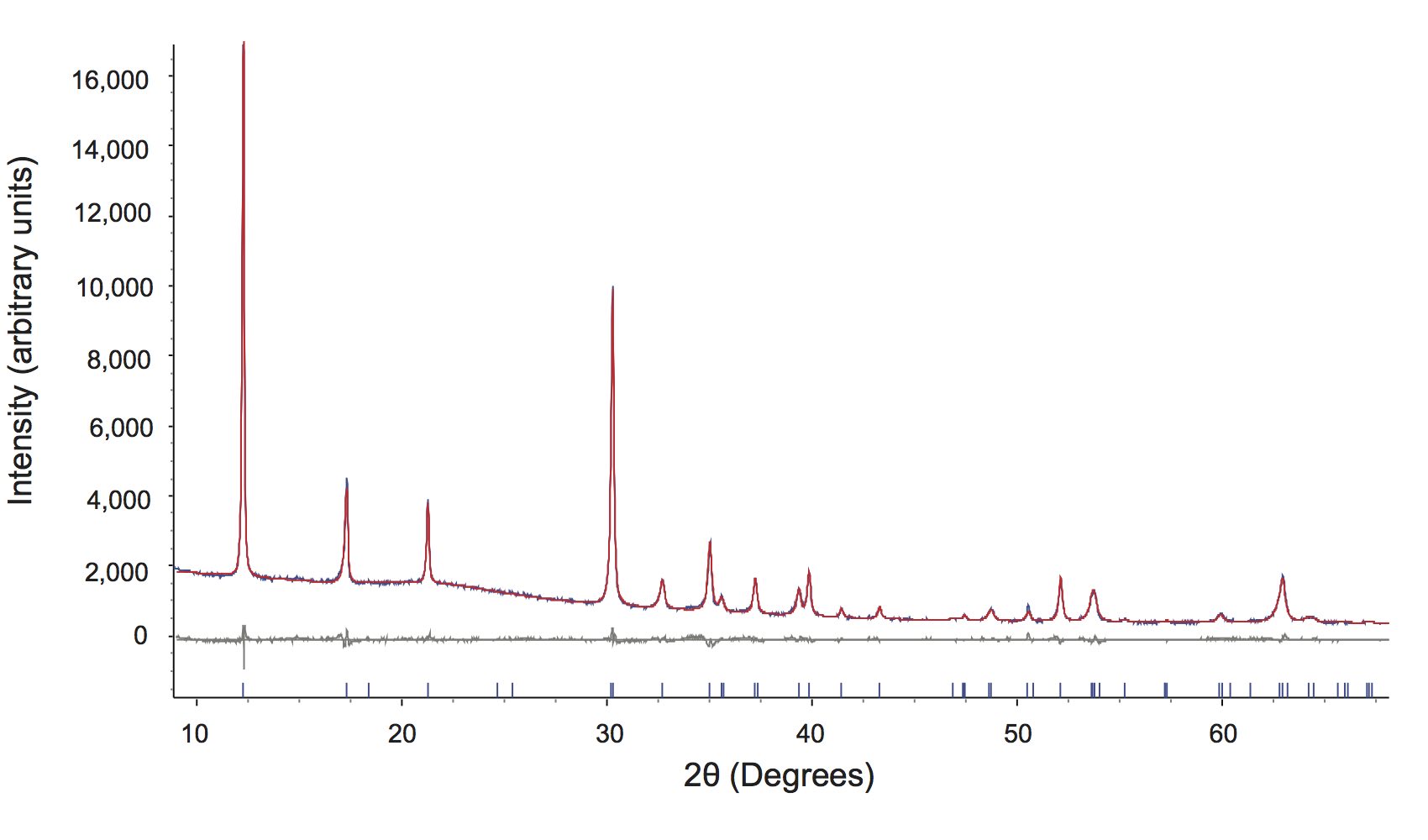}\centering
\caption{Rietveld refinement of XRD data measured on a powdered sample of $\mathrm{NH_4Cu_2.5V_2O_7(OH)_2.H_2O}$ at a wavelength of $\lambda=1.5406\,\mathrm{\AA}$. The red, blue and grey lines, and blue markers represent the fit, data, difference plot and reflection positions, respectively. The final goodness-of-fit parameter was $\chi^2=2.35$ with 56 variables.}
\label{Riet_refine}
\end{figure*}

The powder XRD data was recorded on a STOE Stadi-P diffractometer using Cu-K$\alpha_1$ radiation ($\lambda=1.5406$\,\AA) with a rotating capillary sample holder. As no crystal structure is known for NH$_4$Cu$_{2.5}$V$_2$O$_7$(OH)$_2$.H$_2$O, the starting lattice parameters, space group and atomic positions for the crystal structure model were based on the engelhauptite (KCu$_3$V$_2$O$_7$(OH)$_2$Cl) structure in the $P6_3/mmc$ space group\cite{Pekov2015}. The Rietveld refinement was carried out using the TOPAS software package\cite{Topas}. The data, final calculated and difference plots are shown in Figure \ref{Riet_refine}. The crystal structure data obtained from the refinement is displayed in Table \ref{Crys}; details of the data collection procedure and refinement are given in the supplementary information. All crystal structure figures were produced using VESTA \cite{Vesta}. 

\begin{table}
\caption{\label{Crys}The crystal structure data for $\mathrm{NH_4Cu_2.5V_2O_7(OH)_2.H_2O}$ displaying the atom, Wyckoff site, atomic coordinates, thermal parameter and occupancies.}
\begin{tabular}{@{}lllllll}
\br
Atom & Wyckoff site & $x$ & $y$ & $z$ & B$\mathrm{_{iso}\,(\AA^2)}$ & Occ. \\
\mr
Cu       &       6{\it g}        & $\frac{1}{2}$  &  $\frac{1}{2}$ & 0  &  4.1(11)    & 0.7294(68) \\ 
V        &       4{\it e}         &  0 & 0  &  0.37355(19) &  2.5(11)    & 1 \\
O(1)         &   12{\it k}             &  0.15623(87) & 0.3125(17)  &  0.59043(31) &   3.4(11)   & 1 \\
O(2)     &       2{\it b}         & 0  & 0  & $\frac{1}{4}$  &    5.0(11)  & 1 \\
O(H)     &       4{\it f}         & $\frac{1}{3}$  &  $\frac{2}{3}$ &  0.06520(61) &  2.1(10)    & 1 \\
O(w)         &   6{\it h}             & 0.74537  & 0.3707(98)  &  $\frac{3}{4}$ &   5.8(75)   & 0.242(25) \\
N         &      4{\it f}          & $\frac{1}{3}$  & $\frac{2}{3}$  & 0.2551(19)  &  2.1(20)    & 0.417(21) \\
H(1)         &   4{\it f}             & $\frac{1}{3}$  & $\frac{2}{3}$  &  0.61500 &   5.7 \cite{Lafontaine1990}    & 1 \\ 
H(2)         &  24{\it l}            & 0.64126  & 0.44028  &  0.76082 &    5.7 \cite{Lafontaine1990}   & $=\frac{\mathrm{Ow_{occ}}}{2}$ \\ 
H(3)         &   12{\it k}             & 0.18614  & 0.59307   & 0.74228  &    5.7 \cite{Lafontaine1990}   & $=\mathrm{N_{occ}}$ \\ 
H(4)         &  4{\it f}             &  $\frac{1}{3}$ & $\frac{2}{3}$  & 0.81662  &  5.7 \cite{Lafontaine1990}    & $=\frac{\mathrm{N_{occ}}}{2}$ \\ 
\end{tabular}
\end{table}

\subsection{Structural characterization} \label{SC}
\label{Substr}

NH$_4^+$   and H$_2$O were assigned respectively to the framework cavities occupied by K$^+$ and Cl$^-$ in the similarly structured engelhauptite \cite{Pekov2015}. Structural refinements indicated that these sites feature significant disorder  which was modeled by lowering the symmetry of the N site from $2d$ to $4f$ and the O(w) site from $2c$ to $6h$. To help reveal how hydrogen bonding could stabilise the positions of these species, rigid bodies were defined with the bond lengths and angles set as the following for NH$_4^+$: N\---H(3,4) $=\,0.974\,\mathrm{\AA}$, $\angle$H(3,4)\---N\---H(3,4) = 109.5$^\circ$ \cite{Hawthorne1977}, and for H$_2$O and the OH$^-$ group: O(w)\---H(2) $=\,1.019\,\mathrm{\AA}$, O(H)\---H(1) $=\,1.008\,\mathrm{\AA}$ and $\angle$H(3)\---O(w)\---H(3) = 109.5$^\circ$ \cite{Lafontaine1990}

During the refinement, the site occupancy of the divanadate and hydroxide groups were fixed to be unity while the occupancies of the Cu$^{2+}$, NH$_4^+$  and H$_2$O groups were freely refined. The refined structural formula is (NH$_4$)$_{0.834}$Cu$_{2.188}$V$_2$O$_7$(OH)$_2$.{0.726}(H$_2$O), indicating that some loss of Cu$^{2+}$ occurs that is presumed to be charge compensated by protonation of the water molecules to form hydronium ions. (The latter cannot be distinguished from these data.)

\subsection{Structural analysis}
\label{Structural analysis}

\begin{figure*}[t]
\includegraphics[scale =0.23]{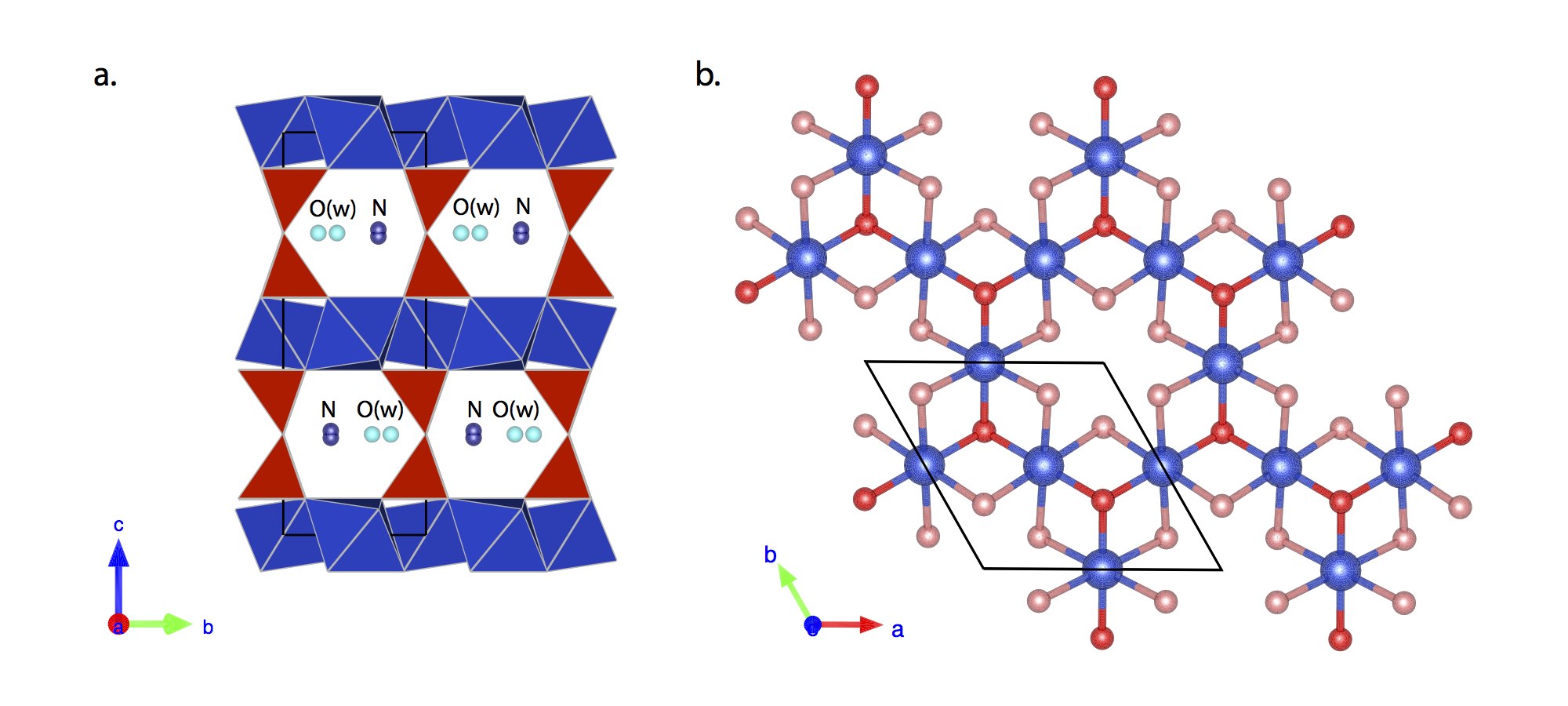}\centering
\caption{\textbf{a.} The structure of $\mathrm{NH_4Cu_2.5V_2O_7(OH)_2.H_2O}$ observed down the $a$-axis. The Cu-octahedra sheets and bivanadate layers are illustrated in blue and red, respectively. The oxygen  of the $\mathrm{H_2O}$ and the nitrogen of the $\mathrm{NH_4^+}$ are shown in the interstitial sites. \textbf{b.} The kagome plane viewed down the $c$-axis. The Cu$^{2+}$ ions (black) sit on a 6$g$ site of the $P6_3/mmc$ space group which has 3-fold rotational symmetry and so the Cu$^{2+}$ ions form a `perfect' kagome. The Cu$^{2+}$ ions ferromagnetic and antiferromagnetic superexchange pathways $via$ the O(2) (red) and O(H) (pink) species, respectively, are shown.}
\label{Struc}
\end{figure*}

Selected bond distances and angles which are pertinent to the description and discussion of the structure of NH$_4$Cu$_{2.5}$V$_2$O$_7$(OH)$_2$.H$_2$O are displayed in Table \ref{Bonds} and the refined structure viewed along the $a$- and $c-$ axes is displayed in Fig \ref{Struc}. The magnetic moments reside on brucite-type Cu-octahedra sheets that are separated by pyrovanadate pillars and interstitial pores, the latter contain NH$_4^+$ and H$_2$O. The interlayer Cu\---Cu separation of 7.22~$\mathrm{\AA}$ is very similar to that of volborthite \cite{Lafontaine1990} (7.21\,\AA) suggesting that the superexchange between layers will be very weak and the magnetic Hamiltonian will be highly 2-dimensional.  The stacking of the kagome layers differs for the 2 materials \-- in volborthite they are in phase while those of  NH$_4$Cu$_{2.5}$V$_2$O$_7$(OH)$_2$.H$_2$O are staggered.

The $\mathrm{Cu^{2+}}$ ions lie on a $2/m$ point symmetry site to form an isotropic kagome lattice, made up of identical equilateral triangular units (Cu\---Cu=2.958~$\mathrm{\AA}$). The isotropic kagome lattice has a ground state Hamiltonian with few exchange terms which better simulates the simple model of theory. 

The Cu on the 6$g$ site was freely refined to an occupancy of $\sim73$\%, a notable deviation from the idealised level that would reduce the number of resonant states available to a QSL state\cite{Lacroix2011}. Despite this, the occupancy is still higher than the bond percolation threshold for a kagome ($p^{\mathrm{bond}}_c$=52\%) \cite{Feng2008}. Further, this situation does not necessarily exclude the possibility of a QSL state as further neighbour entanglement may still cause sufficient degeneracies to allow one to occur. We note the that similar levels of site disorder are seen in some kapellasite samples where they do not to destroy its QSL ground state \cite{Fak2012}.

\begin{table}
\caption{\label{Bonds}Selected bond distances and angles from the structure of NH$_4$Cu$_{2.5}$V$_2$O$_7$(OH)$_2$.H$_2$O}
\begin{indented}
\item[]\begin{tabular}{@{}llll}
\br
\multicolumn{2}{l}{Interatomic distances  ($\mathrm{\AA}$)} & \multicolumn{2}{l}{Angles ($^\circ$)} \\
\mr
Cu-O(H)                           & 1.9373(44)             & Cu-O(H)-Cu          & 99.54(29)       \\
Cu-O(1)                            & 2.1881(43)             & Cu-O(1)-Cu          & 85.05(21)       \\
                                   &                        & O(1)-Cu-O(1)        & 91.63(38)       \\
                                   &                        & O(H)-Cu-O(1)       & 92.48(20)       \\
V-O(2)                             & 1.7867(25)             & V-O(2)-V            & 180.00          \\
V-O(1)                             & 1.6881(78)             & O(2)-V-O(1)         & 108.19(19)      \\
                                   &                        & O(1)-V-O(1)         & 110.72(17)      \\
N-H(4)$\cdots$O(1)                 & 2.182(73)              &                     &                 \\
N-H(3)$\cdots$O(w)                 & 2.060(69)              &                     &                 \\
O(H)-H(1)$\cdots$O(w)             & 1.749(21)              &                     &                
\end{tabular}
\end{indented}
\end{table}

The Cu\---Cu nearest-neighbour superexchange interactions in NH$_4$Cu$_{2.5}$ V$_2$O$_7$(OH)$_2$.H$_2$O are mediated by a O(H) group which sits on a 3-fold axis with a mirror plane, and O(1) which also lies on a mirror plane. The bridging angles of $\angle$ Cu\---O(H)\---Cu = 99.54(29)$^\circ$ and $\angle$ Cu\---O(1)\---Cu = 85.05(21)$^\circ$ are expected to mediate antiferromagnetic and ferromagnetic exchange, respectively, based on the Goodenough-Kanamori-Anderson rules \cite{Goodenough1955, Kanamori1959}. Weak exchange interactions can be expected through both of these pathways as they are close to the cross-over angle at 90$^\circ$ that is a minimum in both ferromagnetic and antiferromagnetic exchange. Such a situation increases the importance of other low-energy exchange terms to the magnetic ground state, illustrated by Dzyaloshinskii-Moriya exchange's  role in forming vesignieite's unique frozen and dynamic spin ground state \cite{Zorko2013}. The Cu\---O bond distances in NH$_4$Cu$_2.5$V$_2$O$_7$(OH)$_2$.H$_2$O  and and the angles involved in superexchange across the kagome lattice are strikingly similar to those of vesignieite, Table \ref{vesig}. As both materials are then expected to display similar magnetic properties, the Dzyaloshinskii-Moriya exchange is also likely to be important here.

\begin{table}
\caption{\label{vesig} Cu\---O bond distances and angles for NH$_4$Cu$_2.5$V$_2$O$_7$(OH)$_2$.H$_2$O and two structures of vesignieite which mediate superexchange are listed for comparison: the crystal structure of vesignieite is disputed so comparisons were drawn with both the monoclinic $C2/m$ \cite{Zhesheng1991} and trigonal $P3_121$ \cite{Boldrin2015b} structures; equivalent bond distances and bond angles from each structure have been compared. Some oxygens in the $P3_121$ structure exhibit disorder over multiple sites; two values have been stated for bond lengths and angles formed with these oxygens.}
\begin{indented}
\item[]\begin{tabular}{@{}llll}
\br
                        & NH$_4$Cu$_2.5$V$_2$O$_7$(OH)$_2$.H$_2$O & BaCu$_3$V$_2$O$_8$(OH)$_2$    & BaCu$_3$V$_2$O$_8$(OH)$_2$\\
												& $P6_3/mmc$															& $C2/m$												&  $P3_121$ \\
\mr
Cu\---O(H)                  & 1.93691(63) \AA                    & 1.913(2) \AA                  &  1.91051(74) \\
Cu\---O                     &  2.18777(84) \AA                         &  2.183(2) \AA            &  2.07250(28) / 2.14803(75) \\
$\angle$ Cu\---O(H)\---Cu & 99.54(29) $^\circ$                          & 101.7(4) $^\circ$       &  103.15(66) \\
$\angle$ Cu\---O\---Cu    & 85.05(21) $^\circ$                          & 85.6(9) $^\circ$        &  92.94(34) / 83.47(54)\\
\end{tabular}
\end{indented}
\end{table}


An important structural feature that is seen in several of the Cu-vanadate kagome magnets \cite{Yoshida2012a, Okamoto2009, Boldrin2015} is an unusual axial compression [2+4] Jahn-Teller distortion\cite{Boldrin2015}, formed of 2 short axial bonds and 4 longer equatorial bonds:  Cu\---O(H)=1.9373(44)\,\AA ~and Cu\---O(1)= 2.1881(43)\,\AA, respectively. At first sight, this configuration could taken to indicate localisation of the unpaired electron in the $d_{x^2-y^2}$ orbital of Cu$^{2+}$, but it has been argued that this type of distortion in powder diffraction is commonly an artifact of averaging when the half-occupied d$_{z^2}$ orbital fluctuates between two degenerate orientations\cite{Burns1996}. We therefore conclude that the Jahn-Teller effect of the Cu$^{2+}$ ions in NH$_4$Cu$_{2.5}$ V$_2$O$_7$(OH)$_2$.H$_2$O is dynamic at room temperature and that the Cu$^{2+}$ orbitals fluctuate.

\begin{figure*}[t]
\includegraphics[scale=0.28]{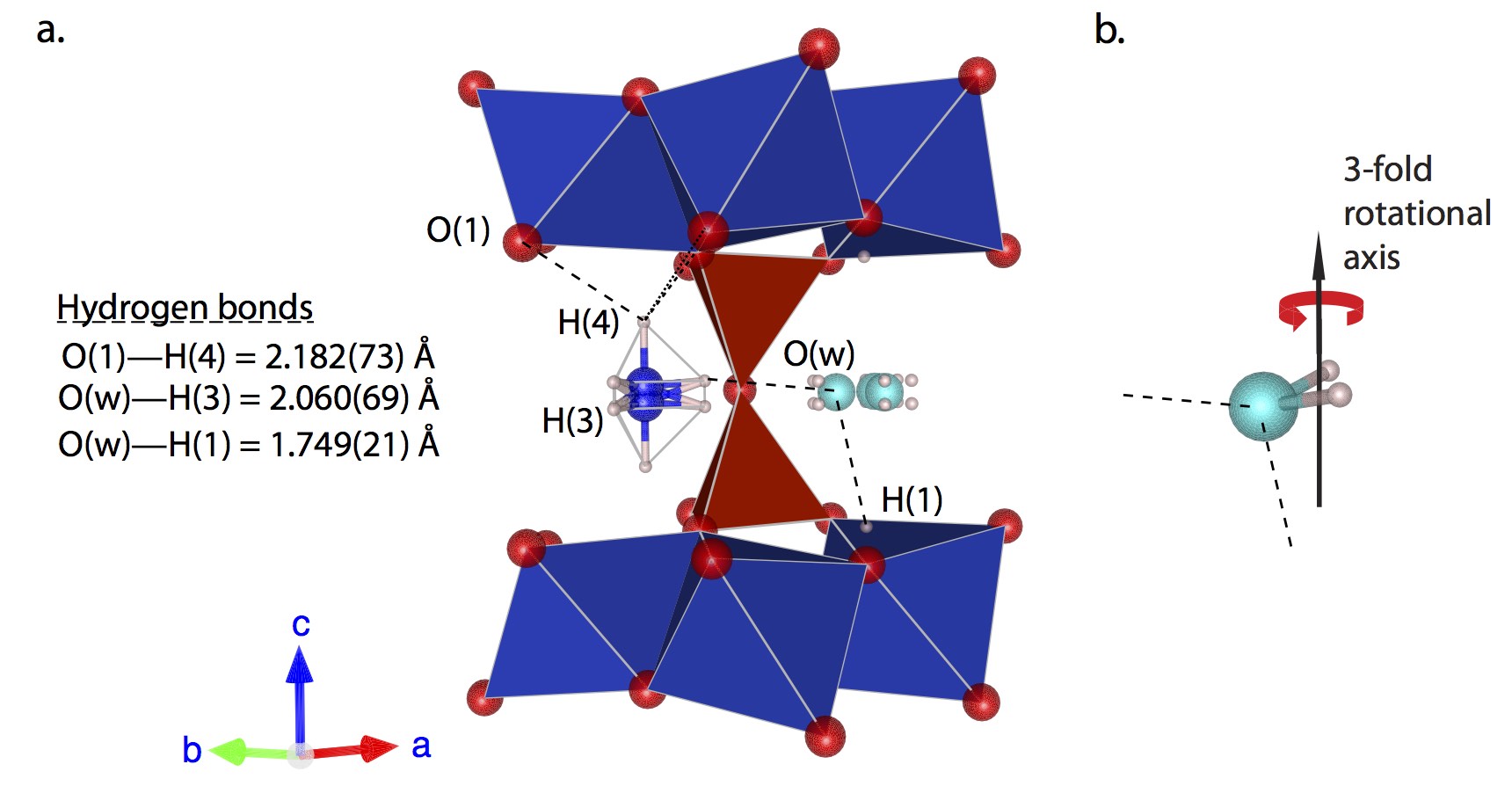}\centering
\caption{\textbf{a} The interstitial NH$_4^+$ and H$_2$O molecules are displayed sitting on the disordered sites between the Cu-octahedra sheets that make up the kagome layers. NH$_4^+$ has two equivalent energy orientations that are mirror images in the $ab$-plane. N-H(4) lies along the $c$-axis and forms three equivalent hydrogen bonds with O(1). The N-H(3) bonds lie on the planes of the 3-fold axis and H(3) forms a hydrogen bond with O(w); O(w) lies off the 3-fold axis and also hydrogen bonds to H(1). \textbf{b} A single H$_2$O rigid body is displayed with the hydrogens canted out of the $ab$-plane as observed in volborthite \cite{Hiroi2001}: the lone pairs of O(w) are now pointing at H(1) for hydrogen bonding.}
\label{Interstial_bonds}
\end{figure*}

Our refinements also show that the NH$_4^+$ unit is displaced from the high symmetry $2d$ site along the $c$-axis; this displacement is likely stabilized by the formation of 3 equivalent N\---H(4)$\cdots$O(1) = 2.182\,$\mathrm{\AA}$ hydrogen bonds and a linear N\---H(3)$\cdots$O(w) = 2.060\,$\mathrm{\AA}$ (Figure \ref{Interstial_bonds}). In turn, displacement of the H$_2$O from the $2c$ site is such that the hydrogens  point below the $ab$-plane of the O(w) site,  as is also observed in volborthite \cite{Hiroi2001}, which would allow the lone pair orbitals of O(w) to point towards H(1), forming O(H)\---H(1)$\cdots$O(w) = 1.987\,$\mathrm{\AA}$. The orientations of these extra-framework molecules could be confirmed through neutron diffraction on a deuterated sample. 

\section{Magnetic characterisation}
\label{Magnetic characterisation}

\begin{figure*}[t]
\includegraphics[scale=0.10]{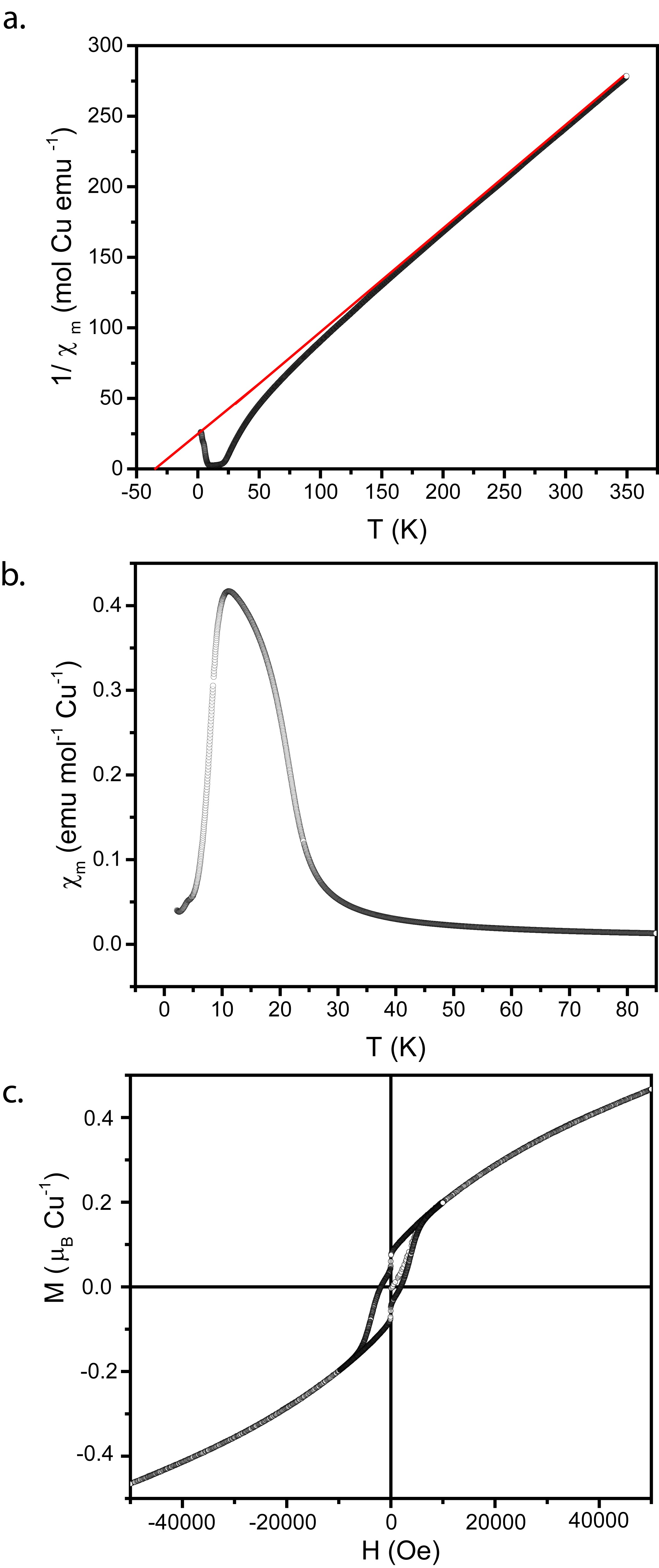}\centering
\caption{All magnetization data was collected on a zero-field cooled sample of NH$_4$Cu$_{2.5}$V$_2$O$_7$(OH)$_2$.H$_2$O \textbf{a.} The $\chi_m^{-1}$ $vs$ $T$ plot shows a deviation from the linear Curie-Weiss law at $T\leq170$\,K due to a build up of local spin correlations. Extrapolation from the linear slope yields a Weiss temperature of $\theta_\mathrm{W} \simeq -30$\,K indicating antiferromagnetic exchange, the absence of an antiferromagnetic transition at $T=30$\,K indicates a frustration of magnetic ordering characteristic of quantum kagome magnets.  \textbf{b.} A plot of $\chi_m$ $vs$ $T$ shows a ferromagnetic transition at $T_{\mathrm{C}} \sim 17$\,K \textbf{c.} $M$ $vs.$ $H$ at 2\,K hysteresis loop with steps, confirming ferromagnetic ordering and indicating a build up of magnetic domains.}
\label{magnetism}
\end{figure*}

Zero-field cooled magnetisation data were collected from NH$_4$Cu$_{2.5}$V$_2$O$_7$(OH)$_2$.H$_2$O (65.3 mg) using the vibrating sample magnetometer of a Quantum Design PPMS-9T in a field of 1000 Oe and with heating rate of 2~K/min. Inspection of $\chi$ {\it vs}. $T$ indicates that there is a ferromagnetic-like transition at $T_{\mathrm{C}}\sim17$\,K that leads to a broad maximum, while the temperature-dependence of the inverse susceptibility shows a linear Curie-Weiss regime over the range $170\leq T \leq 400$\,K from which a Weiss temperature of $\theta_\mathrm{W} \simeq -30\,\mathrm{K}$ can be extrapolated. The deviation from linear behaviour on cooling below 170~K indicates that while spin correlations are building up,  ordering is suppressed to a temperature below $T=|\theta_\mathrm{W}|$: a well known characteristic of $S=\frac{1}{2}$ frustrated magnets\cite{Hiroi2009,Colman2008,Colman2010}. The juxtaposition of a ferromagnetic-like transition and a negative Weiss temperature indicates there is a competition within the magnetism of NH$_4$Cu$_{2.5}$V$_2$O$_7$(OH)$_2$.H$_2$O. The dominant character of the mean field appears to be antiferromagnetic and the transition only occurs when a ferromagnetic energy scale becomes relevant. The room temperature value of the effective moment, $\mu _{\mathrm{eff}} = 2.06~\mu _\mathrm{B}$, is higher than the spin-only value of $\mu _{\mathrm{eff}} = 1.73\, \mu _\mathrm{B}$ indicating that the Land\'{e} g-factor for Cu$^{2+}$ exceeds 2 and that there is an orbital contribution to the magnetism\cite{Zorko2008}.  

Hysteresis in the magnetic field-dependence of the magnetisation data at 2~K shown in Figure \ref{magnetism}c confirms a coherent ferromagnetic component in the low temperature state and an unsaturated paramagnetic signal up to 5~T. The coexistence of ferromagnetic and paramagnetic signals has been observed for some other $S=\frac{1}{2}$ kagome magnets, namely vesignieite, haydeeite, and `Mg-herbertsmthite' \cite{Yoshida2012, Colman2010, Colman2011a}. The contribution of the ferromagnetic and paramagnetic signals to the hysteresis were previously isolated for haydeeite using a paramagnetic Brillouin function with a constant term that accounts for the ferromagnetic response \cite{Boldrin2015b}. In order to extract the ferromagnetic contribution and obtain a value of the spontaneous magnetisation independent of a paramagnetic signal the following equation was used to fit the hysteresis data for NH$_4$Cu$_{2.5}$V$_2$O$_7$(OH)$_2$.H$_2$O:

\begin{eqnarray} \label{2} 
M(H)/M_{sat} = (1 - f)B_{\mathrm{J,PM}}(H) + f     
\end{eqnarray}

\noindent
where M$_{sat}$ is the saturated magnetisation, $f$ is the ferromagnetic response constant and $B_{\mathrm{J,PM}}$ is the paramagnetic Brillouin function per molecule,

\begin{eqnarray} \label{3}
B_{\mathrm{J,PM}}(H)=tanh(g\mu_{\mathrm{B}}JH/k_{\mathrm{B}}T)
\end{eqnarray}

\begin{figure*}[t]
\includegraphics[scale=0.10]{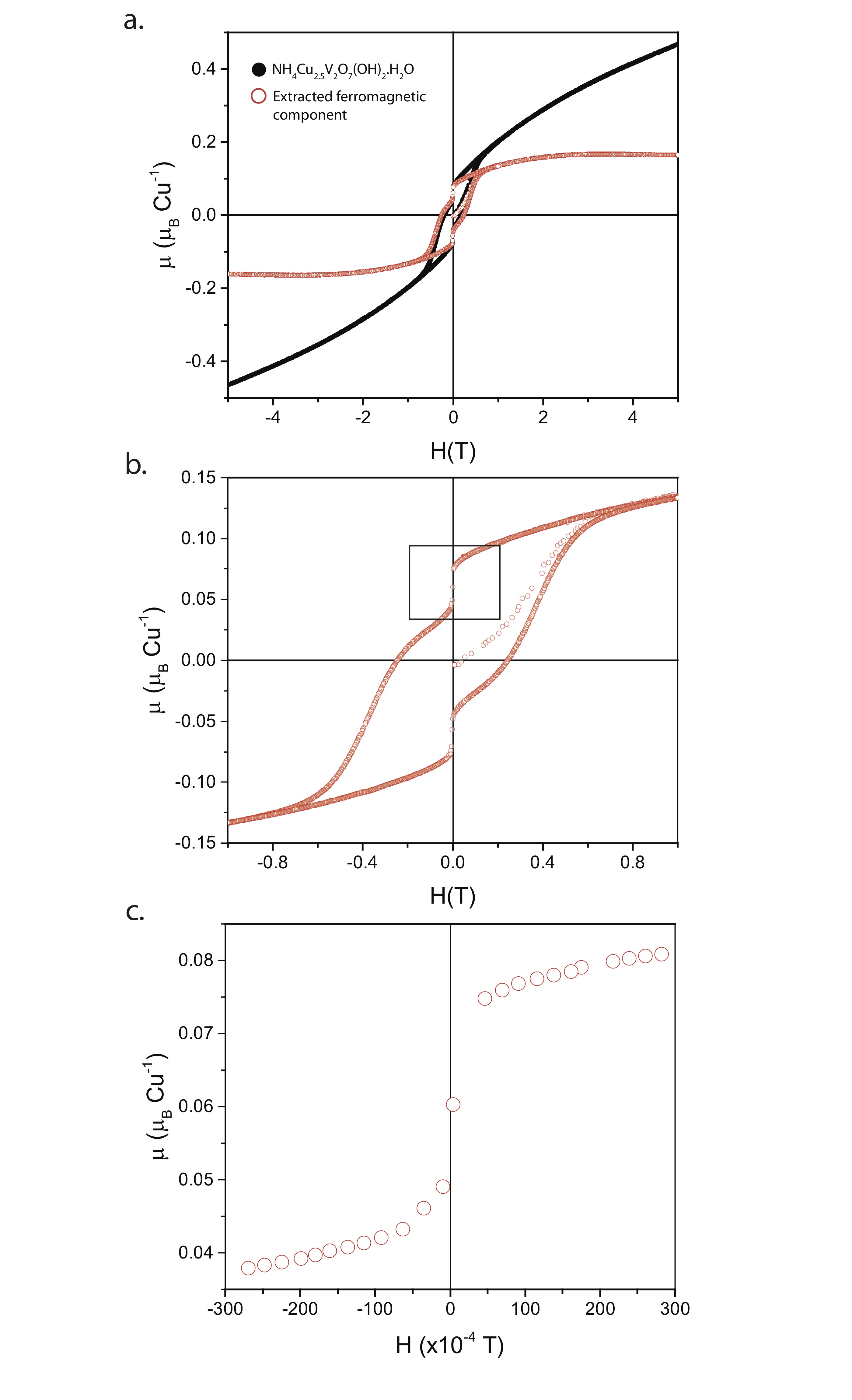}\centering
\caption{\textbf{a} The extracted ferromagnetic contribution (red circles) to the hysteresis is shown along with the $M~vs~H$ data taken at 2\,K  (black circles) \textbf{b} The ferromagnetic hysteresis loop shows  a step near zero-field.  \textbf{c} A close up of region of the step shows it to occur between $-50\,\mathrm{Oe}\lesssim H \lesssim 50$~Oe. }
\label{hyst}
\end{figure*}

\noindent 
Taking Cu$^{2+}$ to be spin only, $J=S=\frac{1}{2}$,  $g$ and $f$ were refined to fit the high-field curve and yielded values of $g=2.36$ and $f=0.41$. This latter value indicates that $\sim41$\% of spins are frozen. A plot of the extracted ferromagnetic contribution to the hysteresis is shown in Figure \ref{hyst}a, and steps in the hysteresis are clearly illustrated in Figure \ref{hyst}b. Magnification of one step, shown in figure \ref{hyst}c, illustrates the sensitivity of the measured moment to small changes of field over the range -50~Oe$\lesssim H \lesssim$50~Oe. This behaviour makes it difficult to determine the spontaneous moment as  zero field coincides with the approximate center of the step, but an upper limit of $\leq 0.075\, \mu_\mathrm{B}\, \mathrm{Cu}^{-1}$ is estimated. The steps in the hysteresis are unusual and are not observed for any other $S=\frac{1}{2}$ kagome magnets,  indicating that the magnetic ground state is indeed exotic and warrants further investigation. 

As discussed in section \ref{Structural analysis}, NH$_4$Cu$_{2.5}$V$_2$O$_7$(OH)$_2$.H$_2$O and vesignieite have very similar superexchange pathways and are expected to display similar magnetic properties. Indeed, both feature antiferromagnetic Weiss temperatures, of  $\theta_\mathrm{W} \simeq -85(5)\,\mathrm{K}$ and $\theta_\mathrm{W} \simeq -30\,\mathrm{K}$ \cite{Colman2011}, respectively, and 
 have magnetic transitions involving a ferromagnetic component \cite{Yoshida2012} to a state with a similar proportion of frozen spins at $T\leq2$~K \cite{Colman2011}. V$^{51}$ NMR and $\mu$SR studies of vesignieite have indicated that the spin ordering  of only partial ($\sim40\,\%$) below $T_{\mathrm{N}}=9$~K and that a dynamic component remains down to 1~K \cite{Colman2011, Quilliam2011}.  ESR analysis of vesignieite revealed that the ordered spin structure was induced by an in-plane Dyaloshinsky-Moriya and is canted out of the kagome plane \cite{Zorko2013}.
 
 A clear difference between the materials is the observed step structure in the hysteresis of  NH$_4$Cu$_{2.5}$V$_2$O$_7$(OH)$_2$.H$_2$O. This may be characteristic of a spin reorientation transition, of the type seen in the metallic  kagome ferromagnetic Fe$_3$Sn$_2$ \cite{Fenner2009}, though here it is unclear whether it would be continuous or involve a $1^{\mathrm{st} }$ order transition and changes in domain occupation.

\section{Conclusion}
\label{Con}

A hydrothermal synthesis and reaction mechanism for the production of pure crystalline samples of the new kagome magnet NH$_4$Cu$_{2.5}$V$_2$O$_7$(OH)$_2$.H$_2$O has been outlined. Its crystal structure consists of `perfect' Cu$^{2+}$ kagome-planes that are separated by V$_2$O$_7^{4-}$ pyrochlore pillars, with ammonium and water groups residing in the interstitial sites; the orientation of the interstitial molecule has been speculated upon. 


Preliminary magnetization measurements indicate  a suppression of magnetic ordering and superparamagnetic correlations that are characteristic of $S=\frac{1}{2}$ kagome antiferromagnets, and a ferromagnetic transition at $T_{\mathrm{C}}\simeq17$\,K. Remarkably, steps are seen in the hysteresis  data at low field which indicate that the ground state has a unusual and marked sensitivity to an applied magnetic field. 

Structural and magnetic similarities between NH$_4$Cu$_{2.5}$V$_2$O$_7$(OH)$_2$.H$_2$O and vesignieite suggest that an in-plane component to the Dyaloshinsky-Moriya exchange plays an important role in spin-ordering. We speculate that this may also involve an orbital freezing transition of the spin-bearing Cu$^{2+}$ ion. Further studies by $\mu$SR  and inelastic neutron scattering studies would help determine the ground state of this unusual kagome magnet and its location in the phase diagram of possible QSLs.

\section{Acknowledgments}

We would like to thank Jeremy Cockcroft for informative discussions, Martin Vickers for experimental assistance, and UCL for the provision of the studentship. 

\clearpage
\section{References}

\bibliographystyle{iopart-num}
\bibliography{NH4Cu2pt5}

\clearpage

\appendix
\begin{appendix}

\section{Supplementary information}

\begin{table}[H]
\caption{\label{Refine}Details of the data collection procedure and Rietveld refinement of $\mathrm{NH_4Cu_2.5V_2O_7(OH)_2.H_2O}$ are displayed along with crystallographic data}
\begin{indented}
\item[]\begin{tabular}{@{}ll}
\br

Crystallographic data       &                      \\
\mr
\hspace{1em} Chemical formula            & NH$_4$Cu$_{2.5}$V$_2$O$_7$(OH)$_2$.H$_2$O \\
\hspace{1em} Crystal system              & Hexagonal                                \\
\hspace{1em} Space group                 & $P6_3/mmc$ (194)                         \\
\hspace{1em} $a$ (\AA)                   &      5.9159(2)                 \\
\hspace{1em} $c$ (\AA)                   &     14.4430(6)              \\
\hspace{1em} Volume ($\mathrm{\AA}^3$)   &         437.76(2)           \\
\hspace{1em} Formula units ($Z$)          &            2          \\
Data collection             &                       \\
\hspace{1em} Radiation, $\lambda$ (\AA)       &        CuK$\alpha$1, 1.54               \\
\hspace{1em} $2\theta$-step size increments ($^\circ$) &       0.05                \\
\hspace{1em} $2\theta$ range ($^\circ$)           &           9 - 70            \\
\hspace{1em} Geometry                    &    Debye-Scherrer geometry            \\
\hspace{1em} Temperature (K)             &              293         \\
\hspace{1em} Zero error ($^\circ$)              &        0.04809(5)               \\
\hspace{1em} $\mu\mathrm{R}$       &    1.6(4)                   \\
\hspace{1em} Number of observed reflections       &              53         \\
Refinement                  &                       \\
\hspace{1em} Instrumental, unit cell and profile parameters        &              36         \\
\hspace{1em} Peak area parameters        &              20         \\
\hspace{1em} Profile function					   &         Stephens' anisotropic broadening \cite{Stephens1999} \\
\hspace{1em} 													   &           Spherical harmonics \cite{Roe1964}\\
\hspace{1em} $\mathrm{R_{exp}}$ \cite{Young2002}     &              1.43          \\
\hspace{1em} $\mathrm{R_{wp}}$  \cite{Young2002}		 &              3.34         \\
\hspace{1em} $\chi ^2$					\cite{Young2002}		 &              2.35         \\
\end{tabular}
\end{indented}
\end{table}
\end{appendix}

\end{document}